\documentclass[a4paper,english,prl,aps,twocolumn, superscriptaddress]{revtex4-2}
\usepackage[T1]{fontenc}
\usepackage{amsmath}
\usepackage{amssymb}
\usepackage{graphicx}

\makeatletter

\pdfpageheight\paperheight
\pdfpagewidth\paperwidth

\makeatletter
\newcommand*{\balancecolsandclearpage}{%
  \close@column@grid
  \cleardoublepage
  \twocolumngrid
}

\makeatother

\usepackage{babel}
\begin{document}
\title{Beyond Constant-Temperature Reservoirs: A Stirling Cycle with Constant
Heat-Generation Rate}
\author{Xinshu Xia}
\address{Graduate School of China Academy of Engineering Physics, Beijing 100193,
China}
\author{Hongbo Huang}
\address{Graduate School of China Academy of Engineering Physics, Beijing 100193,
China}
\author{Hui Dong}
\email{hdong@gscaep.ac.cn}

\address{Graduate School of China Academy of Engineering Physics, Beijing 100193,
China}
\begin{abstract}
Conventional heat-engine models typically assume two heat reservoirs
at fixed temperatures. In contrast, radioisotope power systems introduce
a fundamentally different paradigm in which the hot sources supply
heat at a constant generation rate rather than maintaining a constant
temperature. We develop a theoretical framework for finite-time heat
engines operating between constant heat-generation-rate hot sources
and constant-temperature cold reservoirs. A universal proportion between
average output power and efficiency is established, independent of
the specific cycle configuration or working substance. As a representative
case, we analyze a finite-time Stirling cycle employing a tailored
control protocol that maintains the working substance at constant
temperatures during the quasi-isothermal processes. An intrinsic oscillatory
behavior emerges in the temperature dynamics of the hot source, reflecting
the interplay between heat accumulation and release. We further quantify
the long-term decline in engine performance resulting from radioactive
decay and demonstrate its impact over the system’s operational lifespan.
This work establishes a new theoretical prototype for heat engines
and shall provide guidings for the analysis and design of radioisotope
power systems.
\end{abstract}
\maketitle
A heat engine generally absorbs heat $Q_{\mathrm{h}}$ from a hot
heat reservoir and releases heat $Q_{\mathrm{c}}$ to a cold one,
with the difference converted into mechanical work $W=Q_{\mathrm{h}}-Q_{\mathrm{c}}$.
Its performance is typically evaluated with the efficiency $\eta$
defined as $\eta=W/Q_{\mathrm{h}}$. Under the assumption of the ideal
heat reservoirs with the fixed temperature $\mathcal{T}_{\mathrm{h}}$
and $\mathcal{T}_{\mathrm{c}}$ ($\mathcal{T}_{\mathrm{h}}>\mathcal{T}_{\mathrm{c}}$),
the Carnot efficiency $\eta_{C}=1-\mathcal{T}_{\mathrm{c}}/\mathcal{T}_{\mathrm{h}}$
is proved as the maximum efficiency for any types of engines \citep{huang2008statistical}.
To account for the practical requirement of finite output power, the
finite-time dynamics must be considered \citep{andresen2011current,andresen2022future}.
Earlier attempts for heat reservoirs with fixed temperatures are typically
assumed to reveal the efficiency for heat engines with finite output
power. For example, the efficiency at the maximum power (EMP) $\eta_{\mathrm{CA}}=1-\sqrt{1-\eta_{C}}$
is obtained for the endo-reversible Carnot cycle, where the temperatures
of the working substance in the isothermal processes and heat reservoirs
are both assumed as constant\citep{yvon1955saclay,curzon1975efficiency}.
Contemporary approach with the low-dissipation model goes beyond such
assumption and reveals another EMP as $\eta_{\mathrm{LD}}^{\mathrm{MP}}=\eta_{C}/(2-\eta_{C})$
\citep{esposito2010efficiency}. The existence of these universal
bound for efficiency relies fundamentally on the fact that the temperatures
for both reservoirs are constant during the whole cycle to allow well-defined
Carnot efficiency $\eta_{C}$.

\begin{figure}
\centering{}\includegraphics{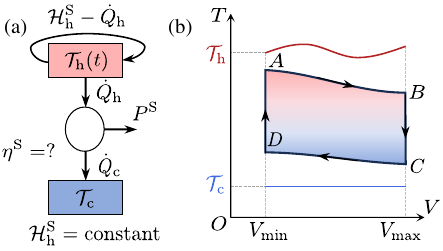}\caption{\label{fig:Stirling_cycle}(a) Schematic diagram of the radioisotope
heat engine with constant heat-generation rate hot source and constant
temperature cold reservoir. The heat engine operates between a hot
source with unfixed temperature $\mathcal{T}_{\text{h}}(t)$ and a
cold reservoir with constant temperature $\mathcal{T}_{\mathrm{c}}$.
The hot source has constant heat-generation rate $\mathcal{H}_{\mathrm{h}}^{\mathrm{S}}$,
of which part $\dot{Q}_{\mathrm{h}}$ outputs heat flow to the heat
engine, and the other part $\mathcal{H}_{\mathrm{h}}^{\mathrm{S}}-\dot{Q}_{\mathrm{h}}(t)$
heats up the hot source with unfixed temperature $\mathcal{T}_{\text{h}}(t)$.
The working substance absorbs heat from the hot source with heat flow
$\dot{Q}_{\mathrm{h}}$, and releases heat to the cold reservoir with
heat flow $\dot{Q}_{\mathrm{c}}$, during which the radioisotope heat
engine extracts work with the output power $P^{\mathrm{S}}$ and efficiency
$\eta^{\mathrm{S}}$. (b) the $T-V$ diagram of the current finite-time
Stirling cycle. The working substance starts at the point $A$ and
goes through four processes as follows, (1) The quasi-isothermal expansion
process $A\rightarrow B$, (2) The isochoric cooling process $B\rightarrow C$
with constant volume $V=V_{\mathrm{max}}$, (3) The quasi-isothermal
compression process $C\rightarrow D$, (4) The isochoric heating process
$D\rightarrow A$ with constant volume $V=V_{\mathrm{min}}$. A progressive
color block represents the temperature change. The temperature of
the hot source $\mathcal{T}_{\text{h}}$ and the cold reservoir $\mathcal{T}_{\text{c}}$
are marked with a red curve and a blue line.}
\end{figure}

Yet, a distinct prototype of heat engines emerges in the context of
radioisotope power systems, such as radioisotope thermoelectric generators
(RTGs) and advanced Stirling radioisotope generators (ASRGs). In these
systems, the hot source does not maintain a constant temperature but
instead supplies heat at a constant generation rate, driven by the
radioactive decay of isotopes such as $^{238}\mathrm{Pu}\mathrm{O}_{2}$,
which has a half-life of 87.7 years. As the operating cycle time is
much shorter than the decay timescale, the heat-generation rate can
be treated as effectively constant, while the temperature of the hot
source varies dynamically within each cycle. Despite the practical
importance of such systems, particularly in space missions, a comprehensive
theoretical framework addressing their thermodynamic performance---especially
in finite-time operation---remains lacking.

In this letter, we establish a model of a finite-time Stirling cycle
operating under the hot source with a constant heat-generation rate,
and find the proportional relation between average output power and
efficiency. To resemble the isothermal process in the traditional
Stirling cycle, we design a control protocol under the ideal gas assumption
to hold the temperature of the working substance constant in the quasi-isothermal
process of the cycle to reveal several analytic results for the efficiency
and the oscillatory behavior of the output power.\textbf{ }On a long-term
scale, we quantitatively demonstrate the decay percentage of the power
for the designed lifespan.

Our model for the radioisotope heat engine is illustrated in Fig.
\ref{fig:Stirling_cycle}(a). The heat-generation rate $\mathcal{H}_{\mathrm{h}}^{\mathrm{S}}$
of the hot source (red square) and the temperature $\mathcal{T}_{\mathrm{c}}$
of the cold reservoir (blue square) stays constant. During the cycle,
part of the heat generation $\dot{Q}_{\mathrm{h}}(t)$ is directed
to the working substance (blank circle), and the rest heat $\mathcal{H}_{\mathrm{h}}^{\mathrm{S}}-\dot{Q}_{\mathrm{h}}(t)$
stays in the hot source inducing the dynamical changes of the heat
source temperature $\mathcal{T}_{\mathrm{h}}(t)$. The working substance
releases heat $\dot{Q}_{\text{c}}(t)$ to the cold reservoir with
the output power $P^{\mathrm{S}}(t)=\dot{Q}_{\mathrm{h}}(t)-\dot{Q}_{\text{c}}(t)-\dot{U}$,
where $\dot{U}$ is the change rate of the working substance's internal
energy. The uniqueness of the current cycle is the constant heat-generation
rate $\mathcal{H}_{\mathrm{h}}^{\mathrm{S}}$, rather than the constant
temperature for the hot source in the traditional engines.

In Fig. \ref{fig:Stirling_cycle}(b), we illustrate the finite-time
Stirling cycle, which is the main type of the radioisotope heat engine
\citep{schreiber2007advanced,richardson2007advanced,thieme2004advanced,schmitz2015modular,chan2007development}.
The four processes in the cycle are (1) the quasi-isothermal expansion
process $A\rightarrow B$, (2) the isochoric cooling process $B\rightarrow C$,
(3) the quasi-isothermal compression process $C\rightarrow D$, and
(4) the isochoric heating process $D\rightarrow A$. To allow the
later discussion, we define $t_{X},X\in\{A,B,C,D\}$ as the moment
of the start points of the four processes marked by the subscript
$i\in\{1,2,3,4\}$ in one cycle with the period $\tau$.

\textit{(1) The quasi-isothermal expansion process}. During this process,
the working substance absorbs heat from the hot source with the heat
flow $\dot{Q}_{1}(t)$ following Newton's law of cooling,
\begin{equation}
\dot{Q}_{1}(t)=-\kappa_{\mathrm{h}}[T_{1}(t)-\mathcal{T}_{\mathrm{h1}}(t)].\label{eq:Q_1}
\end{equation}
where $\kappa_{\mathrm{h}}$ is the heat transfer coefficient between
the working substance and the hot source, and $T_{i}(t)$ is the temperature
of the working substance during the $i$-th process. Apart from the
heat flow to the working substance, the rest of the heat $\mathcal{H}_{\mathrm{h}}^{\mathrm{S}}-\dot{Q}_{1}(t)$
induces the dynamical change $\dot{\mathcal{T}}_{\mathrm{h1}}(t)$
of the hot source as
\begin{equation}
C_{\mathrm{h}}\dot{\mathcal{T}}_{\mathrm{h1}}(t)=\mathcal{H}_{\mathrm{h}}^{\mathrm{S}}-\dot{Q}_{1}(t),\label{eq:T_h1}
\end{equation}
where $C_{\mathrm{h}}$ is the heat capacity of the radioisotope as
the hot source. The volume of the working substance is expanded from
$V_{\mathrm{min}}$ to $V_{\mathrm{max}}$ with output work $W_{1}=\int_{A\rightarrow B}p_{1}(t)dV_{1}(t)$,
where $p_{i}(t)$ and $V_{i}(t)$ are the pressure and volume of the
working substance during the $i$-th process.

\textit{(2)The isochoric cooling process.} During the isochoric process,
the volume of the working substance is fixed $V_{2}(t)=V_{\mathrm{max}}$
with no output work as $W_{2}=0$. The working substance is brought
to contact to the cold reservoir with the heat flow $\dot{Q}_{2}(t)=-\kappa_{\mathrm{c}}[T_{2}(t)-\mathcal{T}_{\mathrm{c}}]$,
which causes the dynamical change of the temperature of the working
substance as 
\begin{equation}
C_{V}\dot{T}_{2}(t)=\dot{Q}_{2}(t),\label{eq:T_2}
\end{equation}
where $C_{V}$ is the constant volume heat capacity of the working
substance, and $\kappa_{\mathrm{c}}$ is the heat transfer coefficient
between the working substance and the cold reservoir. In this process,
the hot source is isolated from the working substance, and heats up
itself by the heat-generation rate $\mathcal{H}_{\mathrm{h}}^{\mathrm{S}}$
as
\begin{equation}
C_{\mathrm{h}}\dot{\mathcal{T}}_{\mathrm{h2}}(t)=\mathcal{H}_{\mathrm{h}}^{\mathrm{S}}.\label{eq:T_h2}
\end{equation}

\textit{(3)The quasi-isothermal compression process.} The working
substance is still in contact with the cold reservoir with releasing
heat
\begin{equation}
\dot{Q}_{3}(t)=-\kappa_{\mathrm{c}}[T_{3}(t)-\mathcal{T}_{\mathrm{c}}].\label{eq:Q_3}
\end{equation}
And the hot source is still isolated with heating up as $C_{\mathrm{h}}\dot{\mathcal{T}}_{\mathrm{h3}}(t)=\mathcal{H}_{\mathrm{h}}^{\mathrm{S}}$.
The volume of the working substance is compressed from $V_{\mathrm{max}}$
to $V_{\mathrm{min}}$ with the output work $W_{3}=\int_{C\rightarrow D}p_{3}(t)dV_{3}(t)$.

\textit{(4)The isochoric heating process}. In this process, the working
substance with the constant volume $V_{\mathrm{min}}$ is brought
back to contact with the radioisotope hot source with the dynamical
change of temperature as 
\begin{equation}
C_{V}\dot{T}_{4}(t)=\dot{Q}_{4}(t),\label{eq:T_4}
\end{equation}
where $\dot{Q}_{4}(t)=-\kappa_{\mathrm{h}}[T_{4}(t)-\mathcal{T}_{\mathrm{h}}(t)]$.
And the temperature of the hot source is changed as 
\begin{equation}
C_{\mathrm{h}}\dot{\mathcal{T}}_{\mathrm{h4}}(t)=\mathcal{H}_{\mathrm{h}}^{\mathrm{S}}-\dot{Q}_{4}(t).\label{eq:T_h4}
\end{equation}
No work is out with $W_{4}=0$. In one cycle of the engine, the total
work is $W=\sum_{i=1}^{4}W_{i}$ .

When the system reaches a cyclic steady state, $\mathcal{T}_{\mathrm{h}}(t)$
is a periodic function with the period $\tau$, namely $\mathcal{T}_{\mathrm{h}}(t+\tau)=\mathcal{T}_{\mathrm{h}}(t)$.
Such condition implies that the total heat injected into the working
substance $Q_{\text{h}}=\int_{A\rightarrow B}\dot{Q}_{1}(t)dt+\int_{D\rightarrow A}\dot{Q}_{4}(t)dt$
is $Q_{\text{h}}=\mathcal{H}_{\mathrm{h}}^{\mathrm{S}}\tau$ for one
cycle. The total work is then obtained as $W=\eta^{\mathrm{S}}Q_{\mathrm{h}}$,
which results in the average output power $\overline{P^{\mathrm{S}}}=W/\tau$
as
\begin{equation}
\overline{P^{\mathrm{S}}}=\eta^{\mathrm{S}}\mathcal{H}_{\mathrm{\mathrm{h}}}^{\mathrm{S}}.\label{eq:P_=0003B7}
\end{equation}
Here, we reach the first properties of the proportional relation between
average output power and efficiency. This proportional relation is
different from other finite-time thermodynamic cycles with fixed hot-source
temperature, where a trade-off relation between power and efficiency
exists \citep{curzon1975efficiency,holubec2016maximum,esposito2010efficiency,ma2018universal}.
To obtain the current relation, we have only used assumptions of no
additional heat outlet via other channel than work extraction and
the periodic condition. And the derivations is independent on the
properties of the working substance. Therefore, we expect such proportional
relation is also applicable to other types of cycles with the hot
source of constant heat-generation rate, e.g., Carnot cycle.

To further demonstrate the feature of the current cycle, we specify
the working substance as ideal gas in the following discussion. In
practical applications, helium gas is a common choice as the working
substance in radioisotope Stirling engines \citep{vanderveer2019stirling}.
Under the temperature and pressure conditions of the radioisotope
Stirling cycle, the helium can be treated as ideal gas \citep{cengel2001thermodynamics}
with the equation of state as $p(t)V(t)=nRT(t)$. Here $n$ is the
amount of substance of the gas and $R$ is the universal gas constant.

To resemble the isothermal process in the original Stirling cycle,
we design a control protocol $V_{i}(t)$ ($i=1,3$) to ensure the
constant temperature $T_{i}$ for the working substance with $\dot{Q}_{i}(t)=p_{i}(t)\dot{V}_{i}(t)$
in the two isothermal processes \citep{chen2021extrapolating,gong2016stochastic}.
For the fixed temperatures $T_{1}(t)=T_{\mathrm{\mathrm{h}}}$ and
$T_{3}(t)=T_{\mathrm{c}}$ of the working substance, the control protocol
is explicitly obtained as
\begin{align}
V_{1}(t) & =V_{\mathrm{min}}\exp(\frac{C_{\text{\ensuremath{\mathrm{h}}}}}{nRT_{\text{\ensuremath{\mathrm{h}}}}}\{[\mathcal{T}_{\text{\ensuremath{\mathrm{h1}}}}(t_{A})-T_{\text{\ensuremath{\mathrm{h}}}}-\frac{\mathcal{H}_{\mathrm{\mathrm{h}}}^{\mathrm{S}}}{\kappa_{\text{\ensuremath{\mathrm{h}}}}}]\nonumber \\
 & \times[1-e^{-\frac{\kappa_{\text{\ensuremath{\mathrm{h}}}}}{C_{\text{\ensuremath{\mathrm{h}}}}}(t-t_{A})}]+\frac{\mathcal{H}_{\mathrm{\mathrm{h}}}^{\mathrm{S}}}{C_{\text{\ensuremath{\mathrm{h}}}}}(t-t_{A})\}),t_{A}\leq t\leq t_{B}\nonumber \\
V_{3}(t) & =V_{\text{max}}\exp[-\frac{K}{\tau_{CD}}(t-t_{C})],t_{C}\leq t\leq t_{D},\label{eq:V_13}
\end{align}
where $K=\ln(V_{\mathrm{max}}/V_{\mathrm{min}})$ is the natural logarithm
of the heat engine compression ratio, and $\tau_{CD}=t_{D}-t_{C}=nRT_{\mathrm{c}}K/[\kappa_{\mathrm{c}}(T_{\text{\ensuremath{\mathrm{c}}}}-\mathcal{T}_{\mathrm{c}})]$
is the duration of the isothermal compression process.

Under the control protocol in Eq.(\ref{eq:V_13}), the output work
$W$ in one cycle is obtained as $W=nRT_{\text{\ensuremath{\mathrm{h}}}}\eta_{\mathrm{eC}}K$,
where $\eta_{\mathrm{eC}}=1-T_{\mathrm{c}}/T_{\text{\ensuremath{\mathrm{h}}}}$
is the endo-reversible Carnot efficiency, which is the Carnot efficiency
for the working substances with the fixed temperatures $T_{\mathrm{\mathrm{h}}}$
and $T_{\mathrm{c}}$ in the two isothermal processes. The derivation
of the total work is presented in the end matter. And the efficiency
$\eta^{\mathrm{S}}$ is in turn obtained as
\begin{equation}
\eta^{\mathrm{S}}=\frac{\eta_{\mathrm{eC}}}{1+f\eta_{\mathrm{eC}}/(2K)},\label{eq:efficiency}
\end{equation}
where $f=2C_{V}/nR$ is the number of the thermally accessible degrees
of freedoms, which is a constant for the ideal gas, e.g., $f=3$ for
the helium gas. The current efficiency $\eta^{\mathrm{S}}$ is related
to the properties of the working substance only via the thermally
accessible degrees of freedoms $f$. To increase the efficiency, we
can reduce the thermally accessible degrees of freedoms $f$, e.g.,
$f=3$ for the single-atom gas.

Another important properties of the current cycle is the period $\tau$,
which characterizes the oscillatory behavior of the output power over
time. The smaller the period, the more stable the energy output. Under
such control protocol, the period $\tau$ of the cycle is obtained
as

\begin{equation}
\tau=\frac{nRT_{\mathrm{\mathrm{h}}}}{\mathcal{H}_{\mathrm{h}}^{\mathrm{S}}}(K+\frac{f\eta_{\mathrm{eC}}}{2}),\label{eq:period}
\end{equation}
which is proportional to the amount of substance $n$ for the two
fixed temperature $T_{\mathrm{\mathrm{h}}}$ and $T_{\mathrm{\mathrm{c}}}$
of the working substance during the two isothermal processes. With
the relation in Eq. \ref{eq:period}, we would expect the small amount
of substance $n$ to ensure the stable energy output in the practical
applications.

Taking the advanced Stirling radioisotope generator (ASRG) \citep{lee2015radioisotope}
designing for outer space as an example, we substitute the actual
data for numerical calculation below. Here we have $\mathcal{T}_{\mathrm{c}}=2.7\text{K}$
for the temperature in the outer space, $\mathcal{H}_{\mathrm{h}}^{\mathrm{S}}=1580\text{W}$
\citep{vanderveer2019stirling} for the heat-generation rate, $K=1$
for the natural logarithm of the compression ratio and $m_{^{238}\mathrm{Pu}\mathrm{O}_{2}}=3.2\text{kg}$
for the mass of the nuclear heat source $^{238}\mathrm{Pu}\mathrm{O}_{2}$
\citep{chan2007development}. The highest temperature of the working
substance $T_{\mathrm{h}}$ is typically chosen to make the maximum
temperature of the hot source as $\mathcal{T}_{\mathrm{hmax}}=1100\text{K}$,
considering the material of the heat engine \citep{schmitz2015modular}.
And the lowest temperature of the working substance is $T_{\mathrm{c}}=400\text{K}$
\citep{schmitz2015modular}. We use a cylindrical heat engine with
the radius of $0.3\text{m}$ and a height of $0.4\text{m}$ at a heat
transfer coefficient of $\kappa_{\mathrm{h}}=\kappa_{\mathrm{c}}=8.6\text{W/K}$
\citep{EngineeringToolBox2003}. The helium gas as the working substance
is encapsulated into the heat engine at the temperature $300\text{K}$
and atmospheric pressure $p_{0}=1.013\times10^{5}\text{Pa}$, with
an amount of substance as $n=4.6\text{mol}$.

\begin{figure}
\begin{centering}
\includegraphics{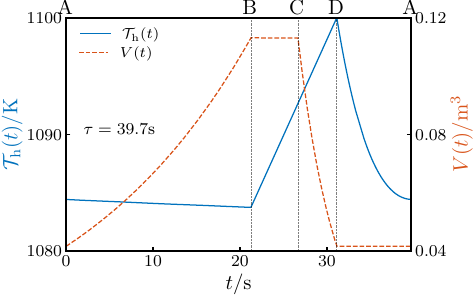}
\par\end{centering}
\caption{\label{fig:The_control_protocol} The variation of the hot source
temperature $\mathcal{T}_{\text{\ensuremath{\mathrm{h}}}}(t)$ and
the control protocol $V(t)$ for one cycle of the finite-time Stirling
cycle. The temperature $\mathcal{T}_{\text{\ensuremath{\mathrm{h}}}}$
of the hot source is plotted with the solid blue line, and the volume
of the working substance $V$ is in the dashed orange line. The period
of the current finite-time Stirling cycle is $\tau=39.7\text{s}$
for the helium gas as working substance.}
\end{figure}

In Fig. \ref{fig:The_control_protocol}, we show the temperature variation
$\mathcal{T}_{\text{\ensuremath{\mathrm{h}}}}(t)$ (solid blue line)
of the hot source over time and the control protocol $V(t)$ (the
dotted orange line) that keeps the temperatures of the working substance
constant for the isothermal expansion and compression processes in
the finite-time Stirling cycle. In the two isothermal processes $A\rightarrow B$
and $C\rightarrow D$, the volume $V(t)$ is plotted as that in Eq.
(\ref{eq:V_13}), and $V(t)$ stays constant in the isochoric processes
$D\rightarrow A$ and $B\rightarrow C$. The period of the current
finite-time Stirling cycle is $\tau=39.7\text{s}$ with the parameters
above.

In the process $D\rightarrow A\rightarrow B$, the heat flow from
the hot source to the working substance exceeds the rate of nuclear
decay, causing a decrease in the temperature of the hot source; in
the process $B\rightarrow C\rightarrow D$, the hot source undergoes
isolated self-heating. The magnitude of the temperature range of the
hot source can be measured by a relative change $\lambda=\Delta\mathcal{T}_{\mathrm{h}}/\mathcal{T}_{\mathrm{hmax}}$,
where $\Delta\mathcal{T}_{\mathrm{h}}=\mathcal{T}_{\mathrm{hmax}}-\mathcal{T}_{\mathrm{hmin}}$
is the amplitude of the temperature fluctuation of the hot source
denoted as the hot source temperature range, and $\mathcal{T}_{\mathrm{hmin}}$
is the minimum temperature of the hot source. Here $\mathcal{T}_{\mathrm{hmax}}=1100\text{K}$
and $\mathcal{T}_{\mathrm{hmin}}=1083\text{K}$, which gives $\lambda=1.5\%$.

\begin{figure}
\begin{centering}
\includegraphics{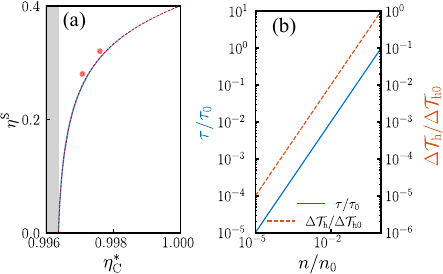}
\par\end{centering}
\caption{\label{fig:Relationship_between_efficiencies}(a) The relation between
the finite-time Stirling cycle efficiency $\eta^{\mathrm{S}}$ and
the average Carnot efficiency $\eta_{C}^{*}=1-\mathcal{T}_{\text{c}}/\overline{\mathcal{T}_{\text{h}}}$.
In the red dashed line, the way $\eta^{\mathrm{S}}$ changes with
$\eta_{C}^{*}$ is shown under constant temperature $T_{\mathrm{c}}$.
For the efficiency $\eta_{C}(t)=1-\mathcal{T}_{\mathrm{c}}/\mathcal{T}_{\mathrm{h}}(t)$
during one cycle, we plot the set of all the curves of the relation
between $\eta^{\mathrm{S}}$ and $\eta_{C}(t)$ as the blue shadow.
The gray forbidden area means $\mathcal{T}_{\mathrm{hmax}}<T_{\mathrm{c}}$,
in which no finite-time Stirling cycle can be constructed. The two
orange points over the blue area are the datas of ASRG\citep{schreiber2007advanced},
which aligns well with our data. (b) The log-log variation of the
period $\tau$ and the hot source temperature range $\Delta\mathcal{T}_{\mathrm{h}}$.
Using the working substance at the standard temperature and atmospheric
pressure with the amount of substance $n_{0}$ as a reference, the
period $\tau$ of the finite-time Stirling cycle with reference $\tau_{0}$
is plotted with the solid blue line, and the temperature range of
the hot source $\Delta\mathcal{T}_{\mathrm{h}}$ with reference $\Delta\mathcal{T}_{\mathrm{h0}}$
is in the dashed orange line.}
\end{figure}

To evaluate the efficiency $\eta^{\text{S}}$, we plot its dependence
on the average Carnot efficiency $\eta_{C}^{*}=1-\mathcal{T}_{\text{c}}/\overline{\mathcal{T}_{\text{h}}}$
defined via the temperature $\mathcal{T}_{\mathrm{c}}$ of the cold
reservoir and the average temperature $\overline{\mathcal{T}_{\text{h}}}=\int_{0}^{\tau}\mathcal{T}_{\text{h}}(t)dt/\tau$
of the hot source with the red dashed line in Fig. \ref{fig:Relationship_between_efficiencies}(a).
Here we fix the temperature $T_{\mathrm{c}}$ of the working substance
in the isothermal compression process, and change $T_{\mathrm{h}}$
in the isothermal expansion process. The blue shadow shows the range
$[\min(\eta_{C}(t)),\max(\eta_{C}(t))]$ of the efficiency $\eta_{C}(t)=1-\mathcal{T}_{\mathrm{c}}/\mathcal{T}_{\mathrm{h}}(t)$
during one cycle. The gray area shows the forbidden region with $\mathcal{T}_{\mathrm{hmax}}<T_{\mathrm{c}}$,
where no heat engine cycle can be constructed. We plot the data from
practical ASRG \citep{schreiber2007advanced} with two orange points
in \ref{fig:Relationship_between_efficiencies}(a). The figure illustrates
that the efficiency predicted in our model is in good agreement with
that of practical ASRG.

In the traditional finite-time thermodynamical cycle with constant-temperature
reservoirs, the performance is typically evaluated with the efficiency
at the maximum power as $\eta_{\mathrm{CA}}=1-\sqrt{1-\eta_{C}^{*}}$
\citep{curzon1975efficiency} or $\eta_{\mathrm{LD}}^{\mathrm{MP}}=\eta_{C}^{*}/(2-\eta_{C}^{*})$
\citep{esposito2010efficiency}. These efficiencies $\eta_{\mathrm{CA}}\thicksim0.94$
($\eta_{\mathrm{LD}}^{\mathrm{MP}}\thicksim0.99$) are clearly higher
than the efficiency obtained for the current cycle with constant heat-generation
rate $\mathcal{H}_{\mathrm{h}}^{\mathrm{S}}$. Hence, the current
prototype of heat engines is significantly different from the traditional
finite-time cycles.

\begin{figure}
\begin{centering}
\includegraphics{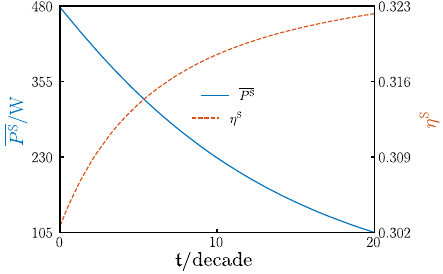}
\par\end{centering}
\caption{\label{fig:decay} The variation of the average output power $\overline{P^{\text{S}}}$
and the efficiency of the finite-time Stirling cycle $\eta^{\text{S}}$
over long-term scales in 20 decades. The the average output power
$\overline{P^{\text{S}}}$ is plotted with the solid blue line, and
the finite-time Stirling cycle efficiency $\eta^{\text{S}}$ is in
the dashed orange line. Since the radioisotope $^{238}\mathrm{Pu}\mathrm{O}_{2}$
decays with half-life period $\tau_{1/2}=8.77$decades, the average
output power $\overline{P^{\mathrm{S}}}$ and the efficiency $\eta^{\mathrm{S}}$
of the cycle also change with time.}
\end{figure}

The period $\tau$ characterizes the oscillatory behavior of the output
power. In Fig.\ref{fig:Relationship_between_efficiencies}(b), we
regulate the amount of substance $n$ and plot the variation of the
period $\tau$ with the solid blue line and the hot source temperature
range $\Delta\mathcal{T}_{\mathrm{h}}$ with the orange dashed line.
The parameters of the cycle at the temperature $300\mathrm{K}$ and
atmospheric pressure by the subscript $0$ as a base for the log-log
plot, such as $n_{0},\tau_{0}$ and $\Delta\mathcal{T}_{\mathrm{h0}}$.
As the Fig.\ref{fig:Relationship_between_efficiencies}(b) shows,
$\tau$ and $\Delta\mathcal{T}_{\mathrm{h}}$ are approximately proportional
to the amount of substance $n$. Additionally, the hot source temperature
range $\Delta\mathcal{T}_{\mathrm{h}}$ increases with $n$. For the
low-density Helium, the hot source with the constant heat generation
rate can be treated as one with the constant temperature.

The uniqueness of the current cycle is its constant heat-generation
over short time. However, such heat-generation rate decays on long-term
scales of years due to the decay of nuclear matter. For example, the
half-life of $^{238}\mathrm{Pu}\mathrm{O}_{2}$ is $\tau_{1/2}=8.77$decades.
To account for the long-term properties, We set $\mathcal{H}_{\mathrm{h}}^{\mathrm{S}}$
as exponential decay $\mathcal{H}_{\mathrm{h}}^{\mathrm{S}}(\mathfrak{t})=\mathcal{H}_{\mathrm{h}}^{\mathrm{S}}(0)\exp(-\mathfrak{t}/\tau_{1/2}\ln2)$,
where $\mathfrak{t}$ represents large-scale time. To allow the proper
comparisons, we fix the maximum temperature of the hot source at $\mathcal{T}_{\mathrm{hmax}}=1100\text{K}$,
and evaluate the power and efficiency of the heat engine. Under this
assumption, we plot the change of the average output power $\overline{P^{\text{S}}}(\mathfrak{t})$
(blue solid line) and the efficiency $\eta^{\text{S}}(\mathfrak{t})$
(the orange dotted line) of the Stirling cycle over time in Fig.\ref{fig:decay}.
In the typical application of ASRG, the designed lifespan of the devices
is $\mathfrak{t}_{\text{life}}=2$decades \citep{vanderveer2019stirling}.
And our work predicts that the relative decay value of the average
output power is $\overline{P^{\text{S}}}(\mathfrak{\mathfrak{t}_{\text{life}}})/\overline{P^{\text{S}}}(0)=87\%$.
To maintain the maximum temperature of the hot source, we has increased
the temperature $T_{\mathrm{h}}$ of the working substance over long-term
scales, which causes the efficiency increases from $\eta^{\text{S}}(0)=0.303$
to $\eta^{\text{S}}(\mathfrak{\mathfrak{t}_{\text{life}}})=0.309$.

In summary, we develop a finite-time Stirling engine model driven
by a hot source with a constant heat-generation rate---a scenario
characteristic of radioisotope-based energy systems, yet largely unexplored
within the framework of finite-time thermodynamics. We establish a
general proportion between average power output and efficiency, which
holds irrespective of the specific cycle implementation or the detailed
properties of the working substance. For the case of ideal gas, we
construct an explicit control protocol that maintains thermal equilibrium
during quasi-isothermal processes, enabling closed-form expressions
for both efficiency and the time-dependent, oscillatory nature of
the power output. When applied to realistic parameter regimes, our
model yields efficiency values in close agreement with those reported
for advanced Stirling radioisotope generators, validating its relevance
to practical systems. We incorporate the decay dynamics of the hot
source and demonstrate how the gradual reduction in heat-generation
rate influences both the power output and thermodynamic efficiency
over extended operational timescales. These results not only extend
the theory of finite-time thermodynamics to a new class of non-isothermal
sources but also provide design principles for long-term power systems
based on radioactive heat.
\begin{acknowledgments}
This work is supported by the Innovation Program for Quantum Science
and Technology (Grant No. 2023ZD0300700), and the National Natural
Science Foundation of China (Grant Nos. U2230203, U2330401, and 12088101).
\end{acknowledgments}

\bibliographystyle{apsrev4-2}
\bibliography{reference}

\balancecolsandclearpage

\section*{End Matter}

\textit{Appendix A: The control protocol of the finite-time Stirling
cycle} --- In the finite-time Stirling cycle, the control protocol
$V_{1,3}(t)$ is designed to fix the temperature of the working substance
in the two quasi-isothermal processes. Combined with the ideal gas
assumption, the working substance has constant internal energy under
the control protocol, leading to the equilibrate between the heat
exchange rate $\dot{Q}_{1,3}(t)$ and work rate $\dot{W}_{1,3}(t)$,
namely $\dot{Q}_{1,3}(t)=p_{1,3}(t)\dot{V}_{1,3}(t)$.

In the quasi-isothermal expansion process during the time $t_{A}\leq t\leq t_{B}$,
the working substance is in contact with the hot source. Combining
with Eq.(\ref{eq:Q_1}), we obtain the equation for the control protocol
$V_{1}(t)$ as $\kappa_{\mathrm{h}}[T_{\mathrm{h}}-\mathcal{T}_{\mathrm{h1}}(t)]+nRT_{\mathrm{h}}\dot{V}_{1}(t)/V_{1}(t)=0$,
with the solution
\begin{equation}
V_{1}(t)=V_{\text{min}}\exp\{\frac{\kappa_{\text{h}}}{nRT_{\text{h}}}[\int_{t_{A}}^{t-t_{A}}\mathcal{T}_{\mathrm{h1}}(\xi)d\xi-T_{\text{h}}(t-t_{A})]\}.\label{eq:V_1_control}
\end{equation}
Despite of the heat directed to the work substance, the additional
heat-generation from the hot source induces the dynamic change the
temperature of the hot source. With Eq.(\ref{eq:Q_1}) and Eq. (\ref{eq:T_h1}),
we obtain the dynamic equation of $\mathcal{T}_{\mathrm{h1}}(t)$
as $C_{\mathrm{h}}\dot{\mathcal{T}}_{\mathrm{h1}}(t)+\kappa_{\mathrm{h}}\mathcal{T}_{\mathrm{h1}}(t)=\mathcal{H}_{\mathrm{h}}^{\mathrm{S}}+\kappa_{\mathrm{h}}T_{\mathrm{h}}(t)$,
which gives the solution as
\begin{equation}
\mathcal{T}_{\mathrm{h1}}(t)=[\mathcal{T}_{\mathrm{h1}}(t_{A})-T_{\text{h}}-\frac{P_{\text{h}}^{S}}{\kappa_{\text{h}}}]\exp[-\frac{\kappa_{\text{h}}}{C_{\text{h}}}(t-t_{A})]+T_{\text{h}}+\frac{P_{\text{h}}^{S}}{\kappa_{\text{h}}}.\label{eq:T_h1_control}
\end{equation}
With Eq.(\ref{eq:V_1_control}) and (\ref{eq:T_h1_control}), we obtain
the first equation in Eq.(\ref{eq:V_13}).

In the quasi-isothermal compression process during the time $t_{C}\leq t\leq t_{D}$,
the working substance is in contact with the cold reservoir with constant
temperature $\mathcal{T}_{\mathrm{c}}$. Combining with Eq.(\ref{eq:Q_3}),
we obtain the equation for the control protocol $V_{3}(t)$ as $\kappa_{\mathrm{c}}(T_{\text{c}}-\mathcal{T}_{\mathrm{c}})+nRT_{\text{c}}\dot{V}_{3}(t)/V_{3}(t)=0$,
with the solution as the second equation in Eq.(\ref{eq:V_13}).

\textit{Appendix B: The efficiency and period of the finite-time Stirling
cycle under the control protocol} --- To show the performance of
the current model with constant heat generation analytically, we use
the ideal gas assumption in the following derivations.

\textit{(1) The quasi-isothermal expansion process}. As the working
substance has constant temperature $T_{\mathrm{h}}$ under the control
protocol, the output work is
\begin{equation}
W_{1}=\int_{V_{\mathrm{min}}}^{V_{\mathrm{max}}}\frac{nRT_{\mathrm{h}}}{V_{1}}dV_{1}=nRT_{\text{h}}K.\label{eq:W_1}
\end{equation}
Constant temperature causes unchanging internal energy for the ideal
gas, which means the heat absorption $Q_{1}=\int_{A\rightarrow B}\dot{Q}_{1}(t)dt$
is all used to extracts work. Namely, $Q_{1}=W_{1}=nRT_{\text{h}}K$.

\textit{(2)The isochoric cooling process.} The volume of the working
substance is fixed as $V_{2}(t)=V_{\mathrm{max}}$ with no output
work $W_{2}=0$. Hence, the heat release $Q_{2}=\int_{B\rightarrow C}\dot{Q}_{2}(t)dt$
is supplied by the internal energy variation as $Q_{2}=\Delta U_{2}=-C_{V}T_{\mathrm{h}}\eta_{\mathrm{eC}}$
of the working substance.

\textit{(3)The quasi-isothermal compression process.} Similar to the
quasi-isothermal expansion process, the output work is
\begin{equation}
W_{3}=\int_{V_{\mathrm{max}}}^{V_{\mathrm{min}}}\frac{nRT_{\mathrm{c}}}{V_{3}}dV_{3}=-nRT_{\text{c}}K.\label{eq:W_3}
\end{equation}
The work done to the working substance is released as heat $Q_{3}$
to the cold reservoir, namely, $Q_{3}=W_{3}=-nRT_{\text{c}}K$.

\textit{(4)The isochoric heating process}. The volume of the working
substance is fixed as $V_{4}(t)=V_{\mathrm{max}}$ with no output
work $W_{4}=0$. Similar to the isochoric cooling process, the heat
absorption $Q_{4}=\int_{D\rightarrow A}\dot{Q}_{4}(t)dt=C_{V}T_{\mathrm{h}}\eta_{\mathrm{eC}}$.

The total output work in one cycle is $W=\sum_{i=1}^{4}W_{i}=nRT_{\text{\ensuremath{\mathrm{h}}}}\eta_{\mathrm{eC}}K$
and the efficiency of the cycle $\eta^{\mathrm{S}}$ at Eq.(\ref{eq:efficiency})
is calculated as
\begin{equation}
\eta^{\mathrm{S}}=\frac{W}{Q_{1}+Q_{4}}=\frac{\eta_{\mathrm{eC}}}{1+f\eta_{\mathrm{eC}}/(2K)}.\label{eq:efficiency_control}
\end{equation}
 And the cycle period is obtained via the power-efficiency relation
in Eq.(\ref{eq:P_=0003B7}) as
\begin{equation}
\tau=\frac{W}{\eta^{\mathrm{S}}\mathcal{H}_{\mathrm{\mathrm{h}}}^{\mathrm{S}}}=\frac{nRT_{\mathrm{\mathrm{h}}}}{\mathcal{H}_{\mathrm{h}}^{\mathrm{S}}}(K+\frac{f\eta_{\mathrm{eC}}}{2}).\label{eq:period_control}
\end{equation}

\end{document}